\documentclass{PoS}
\usepackage{siunitx}
\DeclareSIUnit\fm{\femto\meter}
\DeclareSIUnit\fb{\femto\barn}
\usepackage{amsmath,amssymb}
\usepackage[comma,numbers,sort&compress]{natbib}

\newcommand*\Laplace{\mathop{}\!\mathbin\bigtriangleup}

\newcommand{\lambdabarnew}{{\;\mkern0.75mu\mathchar '26\mkern -9.75mu\lambda}}
\newcommand{\dd}{\ensuremath{\mathrm{d}}}
\newcommand{\potIsoR}{\ensuremath{V_\mathrm{iso}\!\left(r_{4+k}\right)}}
\newcommand{\potIso}{\ensuremath{V_\mathrm{iso}}}
\newcommand{\potr}{\ensuremath{V\!\left(r\right)}}
\newcommand{\GPoisson}{\ensuremath{\bar{G}_D}}
\newcommand{\GBareHigher}{\ensuremath{G_D}}
\newcommand{\horizRad}{\ensuremath{r_\mathrm{h}}}
\newcommand{\lc}{\ensuremath{l_\mathrm{c}}}
\newcommand{\lck}{\ensuremath{l_\mathrm{c}^{\;k}}}
\newcommand{\Rc}{\ensuremath{R_\mathrm{c}}}

\title{Planck scale black holes -- Theory vs.~observations}

\ShortTitle{Planck scale black holes -- Theory vs.~observations}

\author{Michael Florian Wondrak, Piero Nicolini and \speaker{Marcus Bleicher} 
      \\
      
      Frankfurt Institute for Advanced Studies (FIAS)\\
      Ruth-Moufang-Str.~1, 60438 Frankfurt am Main, Germany\\
      
      Institut f\"{u}r Theoretische Physik, Johann Wolfgang Goethe-Universit\"{a}t Frankfurt am Main\\
      Max-von-Laue-Str.~1, 60438 Frankfurt am Main, Germany\\
      
      E-mail: \email{wondrak@fias.uni-frankfurt.de}, 
              \email{nicolini@fias.uni-frankfurt.de}, 
              \email{bleicher@fias.uni-frankfurt.de}}

\abstract{In this paper we present the status of the physics of Planck scale black holes with particular reference to their conjectured production in particle accelerator experiments at the terascale. After reviewing some open issues of fundamental interactions and introducing the physics in the large extra-dimensional scenario, we present the expected signatures left by a microscopic black hole in a particle detector. The final part of the paper is devoted to the latest experimental bounds on the sought black hole signals.}

\FullConference{Frontier Research in Astrophysics -- II\\
		23--28 May 2016\\
		Mondello (Palermo), Italy}

\begin{document}

\section{Introduction}

In a nutshell we can summarize the history of the Universe by saying that after the Big Bang, and its inflationary and reheating phase, the temperature kept dropping inversely proportional to the scale factor, $T\propto a^{-1}(t)$, along a time scale of roughly $\SI{13.8}{Gyr}$. This implies that the Universe underwent phases like nucleosynthesis, recombination and structure formation to gradually approach a situation we observe by looking into the sky today.

Roughly speaking we can say that in a particle accelerator we trace back the above short  history of the Universe, by going in the opposite direction to what naturally occurred. Small amounts of ordinary low-energy matter can be transformed into high-energy states by collisions of particles accelerated to almost the speed of light. For instance at the Large Hadron Collider (LHC), the largest and most powerful particle accelerator ever built,  the Run 2 design center-of-mass energy is  $\sqrt{s}=\SI{14}{\TeV} \approx \SI{2.2e-6}{\J}$, which corresponds to the state of the Universe with temperature $T \approx \SI{1.6e17}{\K}$ or, equivalently, at time $t \sim \SI{e-13}{\s}$ after the Big Bang. The LHC can be helpful to check the Standard Model of particle physics, but more importantly to uncover \textit{new physics} beyond it. We recall here that there are two operational modes at LHC: Colliding protons or heavy-ions, \textit{e.g.}, lead.  In the former case, $\sqrt{s}$ is supposed to reach the aforementioned energy of $\SI{14}{\TeV}$ to look for hints of, for instance, supersymmetry, dark matter, matter-antimatter imbalance as well as unparticle matter. The latter case can be used to produce a quark-gluon plasma and to obtain clues about the stage of the Universe preceding the Big Bang nucleosynthesis. Since also neutrons are involved in a collision, the actual energy per nucleon turns out to be lower: in the case of 
% \ce{^{208}_{82}Pb}, 
${}^{208}_{\,\,\,\,\,82}\mathrm{Pb}$, 
it amounts to about $\SI{2.76}{\TeV}$ \cite{Jowett2008}.

Furthermore, the LHC is able to aim at something else, far more exotic. Based on our everyday experience, we give too often for granted that the world comprises just  three spatial and one temporal dimension. Particle physics itself is formulated on a 4-dimensional flat Minkowski spacetime. We may ask however, if the four dimensional description is  a valid property under any condition or just an approximation that works up to those scales we probed. We recall that  some theories attempting to go beyond the Standard Model even prescribe additional spacelike dimensions. This is the case of string theory and its mature extensions such as Superstring Theory and M-theory, that aim at unifying all fundamental forces including gravity in a 10-dimensional or 11-dimensional spacetime, respectively.
In such a setup our perception can be  guaranteed by the fact that the additional dimensions are compactified. This means that we are restricted to a (3+1)-dimensional slice (called \textit{brane}) of the whole spacetime, conventionally denoted as  \textit{bulk}.

Despite their non-observability in conventional situations, the possibility of additional spatial dimensions may be detectable in current or near future experimental and observational facilities. In this paper we present an overview of the current status of the research in that field.

After a brief review of the status of fundamental interactions  and the hierarchy problem, we focus on the  Arkani-Hamed, Dimopoulos, Dvali (ADD) model of large extra dimensions \cite{Arkani-HamedDD1998,AntoniadisADD1998}. As a potential test for this  model, we describe black holes in higher dimensions and the possibility of their production in particle accelerators \cite{ADM98,BaF99,DiL01,GiT02}. Finally we discuss the experimental results based on the latest data. In the appendices we discuss the compactification of higher dimensions using the gravitational action or the Poisson equation and we study the relations among the various associated gravitational constants.

\section{Gravity, the Hierarchy Problem, and the ADD Model}

The idea of additional dimensions has a long history in physics starting with the theory of Kaluza and Klein in the 1920's \cite{Kaluza1921,Klein1926a,Klein1926b}. From a modern perspective the idea, however, resurfaced at the end of the 1990's, some time before the Run 2 beam energy at the particle accelerator Tevatron reached \SI{980}{\GeV}. 

The premises for additional spatial dimensions are connected to an efficient description of all fundamental interactions. We recall that  the Standard Model has reached a high degree of accuracy and experimental corroboration, 
but though it cannot be considered as a fundamental theory. Apart from open questions related to neutrino masses, the number of external parameters, and possible unifications, the most critical shortcoming is the absence of the gravitational interactions. This fact opens up two additional issues, \textit{i.e.}, the formulation of a quantum theory of gravity and the hierarchy problem. 
 
As is well known, gravity escapes a direct quantization like other interactions due to its non-renormalizable character.  To this purpose one has to assume new quantization schemes or new dynamical properties for quantized objects.  Along such ways of thinking, candidate theories of Quantum Gravity have been formulated, such as Loop Quantum Gravity, Asymptotically Safe Gravity, and Superstring Theory. Common to all of them is the problem of the extraordinary weakness of the gravitational interaction. If we compare the weak and gravitational coupling constants, we find a  ratio
\begin{equation}
\frac{G_{\mathrm{F}}}{G_{\mathrm{N}}} \approx \num{1.7e33},
\end{equation}
where  $G_{\mathrm{F}}$ is Fermi's constant and $G_{\mathrm{N}}$ is Newton's constant. 
Alternatively, we can illustrate the weakness of gravity by evaluating  the relative coupling strength for electromagnetic and gravitational interaction in the classical regime. Coulomb's force $F_{\mathrm{em}}$ and Newton's force $F_{\mathrm{grav}}$ read
\begin{equation}
F_{\mathrm{em}} = \frac{q_1\,q_2}{r^2}, \qquad\qquad
F_{\mathrm{grav}} = G_{\mathrm{N}} \, \frac{m_1\,m_2}{r^2}
\end{equation}
where $q_1$ and $q_2$ denote the electric charges of the interacting particles in Gaussian units, $m_1$ and $m_2$ stand for their masses, and $r$ is the separation distance. Even for the heaviest elementary particle, the top quark with mass $m_\mathrm{top}\approx\SI[per-mode=symbol]{170}{\giga\eV}$,  the smallness of Newton's gravitational constant yields 
\begin{equation}
\frac{F_{\mathrm{em}}}{F_{\mathrm{grav}}} \approx \num{4e31}
\end{equation}
rendering  the gravitational interaction negligible. In evaluating the above ratio we used the fact that $G_{\mathrm{N}} = 1/M_\mathrm{Pl}^2$, where $M_\mathrm{Pl} \approx \SI{1.2e19}{\GeV} \approx \SI{2.2e-8}{kg}$ is the Planck mass, \textit{i.e.}, the scale at which Quantum Gravity sets in. 
This means that the above ratio approaches unity and becomes relevant only if the masses are chosen to be $m_1,\,m_2 \sim M_\mathrm{Pl}$ which is 15 orders of magnitude greater than the highest LHC energies\footnote{The discrepancy between the Planck scale $M_\mathrm{Pl}$ and other fundamental energy scales is called ``hierarchy''.}. In the absence of further hypotheses, the gravitational interaction is actually negligible in particle physics experiments. At first sight, such a weakness does not seem to be a problem apart from the fact that we do not have any explanation for it. On a deeper level, one realizes that the weakness of gravity is rooted with a potential incompleteness of the Standard Model. In other words, we have to expect families of new particles emerging in the energy range from the electroweak scale $\Lambda_{\rm EW}\approx 246$ GeV at least up to the Planck scale. To avoid such a non-converging scenario that would require the formulation of an array of particle models, each for a given energy scale, one has two possible choices. The first one is to deny the problem by assuming a big desert approximation, \textit{i.e.}, the absence of further particles beyond the Standard Model. The other one is to find a mechanism that reduces the energy scale associated with gravity drastically down to the order of the electroweak scale.

By following the latter way, Arkani-Hamed, Dimopoulos, and Dvali together with Antoniadis proposed a possible solution to the hierarchy problem in 1998 by extending the number of dimensions of spacetime \cite{Arkani-HamedDD1998,AntoniadisADD1998,Arkani-HamedDD1999}. Such a proposal which we briefly present below has been accompanied by forerunners \cite{Ant90} and/or competing models such as  the brane-world scenarios, (\textit{e.g.}, the domain wall model \cite{RuS83}, the Randall--Sundrum models \cite{RaS99a,RaS99b}, the  shell Universe model \cite{Gog99,Gog00,Gog02}), and the universal extra dimension scenario \cite{ACD01}.   

While the familiar observable universe lives on a (3+1)-dimensional submanifold (the so-called \textit{D3-brane} plus the time coordinate), the spacetime actually possesses ((3+$k$)+1) dimensions (\textit{bulk})\footnote{In order to preserve causality, the $k$ additional dimensions are spacelike. The bulk coordinates are denoted by $\left(x^0,\,\dots,\,x^3,\,x^4,\,\dots,\,x^{3+k}\right)$. An example of an higher-dimensional flat metric is given by
\begin{equation}
\dd s^2 
= -{\left(\dd x^0\right)}^2 + \sum_{i=1}^3 {\left(\dd x^i\right)}^2 
  + \sum_{i=4}^{3+k} {\left(\dd x^i\right)}^2.
\end{equation}} and is fully accessible only for the gravitational interaction. 
This situation is similar to superstring theory in which gravitons correspond to excitations of closed strings. Only closed strings are supposed to propagate in the additional dimensions. Standard Model particles arise from vibrations of open strings and thus they are restricted to the brane. The number $k$ of extra dimensions is in principle arbitrary. Since superstring theory and M theory are formulated in 9+1 and 10+1 dimensions, respectively, one usually considers up to 7 extra dimensions.

The special treatment of gravitation is the key ingredient to solve the hierarchy problem: A priori, all the four fundamental forces are of similar strength, but gravity affects $k$ more dimensions so that it appears weaker when one just considers the brane. As a result, the new higher-dimensional gravitational constant, which we will call $\GBareHigher$, leads to a much stronger coupling than the ordinary four-dimensional Newton's constant $G_\mathrm{N}$. Analogously, the $D$-dimensional fundamental mass scale for $D=4+k$
\begin{equation}
M_D \equiv {\left( \frac{1}{8\pi}\,\frac{1}{\GBareHigher} \right)}^{1/\left(2+k\right)}
\end{equation}
is much lower than the Planck mass $M_\mathrm{Pl}$. Assuming effects accessible at energies of particle accelerators, the scale is usually set to $M_D \sim \SI{1}{\TeV}$.
The details of the relation between the higher and the 4-dimensional quantities are derived in the appendices \ref{App_Action} and \ref{App_Poisson} with a special focus on the connection with the Kaluza-Klein theory and the Poisson's equation.

The presence of extra dimensions affects the radial profile of the gravitational potential. Experimentally proven for distances above $r^\mathrm{min}_\mathrm{N}=\SI{56}{\micro\m}$, Newton's law dictates a force $F\propto\frac{1}{r^2}$ \cite{AdelbergerGHHS2009}. Therefore the extra dimensions have to be compact with an extension $l_\mathrm{c}=2\pi \Rc$ smaller than $r^\mathrm{min}_\mathrm{N}$. Mathematically, this can be achieved easily by identifying all points on the extra dimension $i$ whose coordinate difference is a multiple of $2\pi {\Rc}_i$. Each of the $k$ extra dimensions may have a different compactification radius ${\Rc}_i$. The compactification of $k$ extra dimensions gives rise to a $k$-torus $T^k$, the topology of spacetime becomes $\mathcal{M}_4 \times T^k$. This means that there is a $k$-torus attached to every event in the submanifold $\mathcal{M}_4$.

For simplicity we will assume in the following that all ${\Rc}_i$ coincide, ${\Rc}_i \equiv \Rc$ for $1 \leq i \leq k$ (symmetric toroidal compactification). The value of $\Rc$ might be much larger than the Planck length in the case of no extra dimension, $l_\mathrm{Pl} \equiv \sqrt{G_\mathrm{N}} = 1/M_\mathrm{Pl} \approx \SI{1.6e-35}{\meter}$ (cf.~\eqref{eq:App_PlMass}).  From \eqref{eq:App_RelationMasses} we obtain the order of magnitude of the compactification radius in the presence of extra dimensions,
\begin{equation}
\Rc 
= \frac{1}{2\pi}\,{\left(\frac{M_\mathrm{Pl}^2}{8\pi\,M_D^{2+k}}\right)}^{1/k} 
\sim 10^{\frac{32}{k}-19}\si{\meter}.
\label{eq:CompactRadius}
\end{equation}
Depending on the number of extra dimensions $k$, one finds $\Rc$  in the range $\Rc \sim \left[ \si{\fm},\,\si{\nano m} \right]$ in order to ensure $M_D \sim \SI{1}{\TeV}$.  For this reason the theory is called \textit{large} extra dimensions.

The careful reader may wonder why the ADD model admits less extra dimensions than superstring theories or M-theory. The two proposals are  compatible since extra dimensions with very small compactification radii can be neglected with respect to extra dimensions with a larger radius. The number of Planckian extra dimensions will not decrease  the value of $M_D$ with respect to $M_\mathrm{Pl}$.

After introducing the concept of extra dimensions one might ask for predictions which can be tested in an experiment. In the absence of extra dimensions, quantum gravity effects are expected at the Planck energy which is 15 orders of magnitude above those accessible by the LHC experiments. In contrast, the ADD model allows detectable effects in a variety of energy scales. At low energies, investigation of the gravitational force at short distances might disclose deviations from inverse-square law indicating the compactification radius $\Rc$ and the number $k$ of extra dimensions; at energies around \si{\MeV} \cite{AdelbergerGHHS2009,ABH07,Hoy04,LHC02}, the cooling processes of supernovae or neutron stars would be modified \cite{Arkani-HamedDD1999}; at energies close to $M_D \sim \SI{1}{\TeV}$, on-shell gravitons could be produced in current particle accelerators \cite{GRW99,GiS03}. The disappearance of gravitons into the bulk would lead to a missing transverse energy $E_\mathrm{T}$. Ultra-high-energy cosmic rays hitting the higher layers of Earth's atmosphere can offer further insights at even higher energies, up to $\sim 10^8$ TeV \cite{EMR02}. The most striking evidence of extra dimensions would be, however, the production of microscopic black holes in particle accelerators. A stronger gravitational interaction allows for the gravitational collapse (in the bulk) of incoming particles flying on the brane. For a list of reviews on the topic see, \textit{e.g.}, \cite{Lan02,Cav03,Kan04,Hos04,CaS06,Ble07,Win07,Nic09,BlN10,BlN14,
Cal10a,Par12,KaW15}.

\section{Black Holes in Higher Dimensions}
General relativity black hole solutions are characterized by the symmetry of spacetime and the three parameters mass, charge, and angular momentum, the so-called black hole hair. According to Birkhoff's theorem, the unique spherically symmetric solution in vacuum is the Schwarzschild geometry.
In 4 dimensions, the Schwarzschild metric is given by
\begin{equation}
\dd s^2
= -\left(1-\frac{2G_\mathrm{N}\,M_\mathrm{BH}}{r}\right) \, \dd t^2
  +\frac{1}{\left(1-\frac{2G_\mathrm{N}\,M_\mathrm{BH}}{r}\right)} \, \dd r^2
  +r^2 \, \dd \Omega_2^2
\end{equation}
If the spatial extension of a spherically symmetric, static object is smaller than the corresponding Schwarzschild radius $\horizRad=2G_\mathrm{N}\,M_\mathrm{BH}$, the object turns out to be a (Schwarzschild) black hole of mass $M_\mathrm{BH}$. The Schwarzschild radius defines the location of the event horizon, \textit{i.e.}, the border which can be crossed by any kind of particle only from the outside.

The concept of a spherically symmetric, static black hole can easily be generalized to 4+$k$ infinitely extended dimensions. In this case, the metric reads \cite{Tangherlini1963,Myers1986}
\begin{equation}
\dd s^2
= -\underbrace{\left(1-\frac{2}{1+k}\,\frac{\GPoisson\,M_\mathrm{BH}}{r^{1+k}}\right)}_{\equiv f\!\left(r\right)} \, \dd t^2
  +\frac{1}{\left(1-\frac{2}{1+k}\,\frac{\GPoisson\,M_\mathrm{BH}}{r^{1+k}}\right)} \, \dd r^2
  +r^2 \, \dd \Omega_{2+k}^2
\end{equation}
where $\GPoisson$ is defined as the gravitational constant appearing in the higher dimensional Poisson's equation \eqref{eq:app_HigherPoisson} and 
\begin{equation}
\dd \Omega_{2+k}^2 \equiv \dd \chi_1^2 
+ \sum_{i=2}^{2+k}\,\prod_{j=1}^{i-1}\, \sin^2\!\chi_j\;\dd \chi_i^2
\end{equation}
is the angular line element in 2+$k$ dimensions. As usual, this metric yields the expected higher-dimensional gravitational potential \eqref{eq:App_IsoPotential} in the weak field limit.

At the black hole horizon, the sign of the metric component $f\!\left(r\right)$ flips indicating the point of no return. Therefore, the horizon's position directly follows from the condition $f\!\left(\horizRad\right)=0$ where
\begin{equation}
\horizRad
=\horizRad\!\left(M_\mathrm{BH},\,k\right)
={\left(\frac{2\,\GPoisson\,M_\mathrm{BH}}{\left(1+k\right)}\right)}^{1/\left(1+k\right)}
% ={\left(\frac{8\,\Gamma\!\left(\frac{3+k}{2}\right)\,\lck\,G_\star\,M_\mathrm{BH}}{\left(2+k\right)\,\pi^{\left(1+k\right)/2}}\right)}^{1/\left(1+k\right)}
={\left(\frac{\Gamma\!\left(\frac{3+k}{2}\right)\,M_\mathrm{BH}}{\left(2+k\right)\,\pi^{\left(3+k\right)/2}\,M_D^{2+k}}\right)}^{1/\left(1+k\right)}.
\end{equation}
Since we are interested in microscopic black holes with mass $M_D\lesssim M_\mathrm{BH} \ll M_\mathrm{Pl}$
the horizon radius $\horizRad$ turns out to be much smaller than the compactification length  $l_\mathrm{c} ={\left(M_\mathrm{Pl}^2/\left({8\pi\,M_D^{2+k}}\right)\right)}^{1/k}$ (cf.~\eqref{eq:CompactRadius}) in the ADD model.
Thus the black hole perceives an isotropic higher dimensional spacetime as in the case of non-compact extra dimensions.\footnote{For a black hole solution in the Randall-Sundrum model RS2 see, \textit{e.g.}, \cite{DadhichMPR2000,AlexeyevS2010}.}

While arbitrarily high masses are allowed, there is, on the other hand, a minimal mass $M^\mathrm{min}_\mathrm{BH}$ for black hole formations in a particle collision: A gravitational collapse occurs if the size of an object falls within its Schwarzschild radius $\horizRad$. This principle can be extended to the small-mass regime where quantum mechanics becomes relevant. The probability distribution determines the smallest size of a quantum object of mass $m$. A typical measure of such a size is the reduced Compton wavelength $\lambdabarnew_\mathrm{C} = \frac{1}{m}$.
Assuming that both concepts of characteristic length scales remain valid we can estimate the characteristics of the smallest possible black hole by requiring that
\begin{equation}
\horizRad \stackrel{!}{=} \lambdabarnew_\mathrm{C}.
\end{equation}
From this one obtains:
\begin{eqnarray}
\begin{array}{l}
M^\mathrm{min}_\mathrm{BH} 
\propto M_D \,
\propto {\left(\dfrac{l_\mathrm{Pl}}{l_\mathrm{c}}\right)}^{k/\left(2+k\right)}\,M_\mathrm{Pl}\\
\\
r^\mathrm{min}_\mathrm{BH}
\propto \dfrac{1}{M_D} \,
\propto {\left(\dfrac{l_\mathrm{c}}{l_\mathrm{Pl}}\right)}^{k/\left(2+k\right)}\,l_\mathrm{Pl}.
\end{array}
\label{eq:BH_min_approx}
\end{eqnarray}
See appendix \ref{App_Mmin_rmin} for the complete formulae.
The above relation is instrumental for the gravity ultraviolet self-completeness also known as classicalization \cite{GRW02,AuS13,DvG10,DFG11,DGG11,DvG12}. This is equivalent to saying that the trans-Planckian collision energies imply higher  masses and bigger horizon radii of the resulting black holes, rather than shorter Compton wavelengths. We notice that classicalization is a purely quantum gravitational feature \cite{DvG13b,DvG13,DvG14,DGI15,CaS14,CMS14,CMN15b,CMOr12,MuN12,
IMN13,FKN16} even if quantum effects are, in general, not sufficient to guarantee the self-completeness \cite{CMN15}. Indeed General Relativity allows for black holes of any mass without a lower bound \cite{NiS12}.  Classically sub-Planckian black holes can exist as a result of a primordial formation for the presence of high density matter fluctuations in early Universe \cite{CaH74} or for a quantum mechanical decay of deSitter space \cite{MaR95,BoH96}. Interestingly sub-Planckian black holes can exist also within some effective approaches to quantum gravity like the Generalized Uncertainty Principle \cite{CMN15}.

In \eqref{eq:BH_min_approx}, if $k$ increases, the minimal mass $M^\mathrm{min}_\mathrm{BH}$ decreases. For properly chosen compactification lengths $\lc$ such a minimal mass is at the reach of particle accelerators. As a result, a black hole could  form if the center-of-mass energy of two colliding objects exceeds $M^\mathrm{min}_\mathrm{BH}$ and their impact parameter $b$ is smaller than the Schwarzschild radius of the effective two-body system, $b < \horizRad$ (cf.~\cite{Thorne1972}). 

\section{Black Hole Formation in an Accelerator}
The production rate of black holes in a collision experiment is encoded in the production cross section. The simplest guess is the black disk approximation, \textit{i.e.}, the cross section equals the geometric cross section corresponding to the Schwarzschild radius, 
\begin{equation}
\hat{\sigma}\!\left(M_\mathrm{BH},\,k\right) 
= \pi\, \horizRad^2\!\left(M_\mathrm{BH},\,k\right)
\end{equation}
(cf.~\cite{Thorne1972}).
In an accelerator, black holes could be produced by the collision of two protons at a center-of-mass energy $\sqrt{s}$. Since protons are not elementary, their substructure has to be considered as the fundamental degrees of freedom: quarks and gluons or, at first order in the strong coupling constant $\alpha_\mathrm{s}$, massless partons. 
A black hole of mass $M_\mathrm{BH}$ could actually form at the collision of parton $a$ from the first proton and parton $b$ from the second proton \cite{DiL01,BleicherHHS2002}.
The cross section for this process becomes
\begin{equation}
f_a\!\left(x_1,\,Q^2\right)\,f_b\!\left(x_2,\,Q^2\right)\,\hat{\sigma}\!\left(\sqrt{\hat{s}},\,k\right).
\end{equation}
Here, the parton distribution function $f_i\!\left(x_j,\,Q^2\right)$ gives the probability to find a parton of type $i$ in the hadron $j$ whose momentum amounts to $x_j$ of the hadron's initial longitudinal momentum, $0 \leq x_j \leq 1$. The additional argument $Q^2$ specifies the exchanged momentum in the process of interest, which can be interpreted as a means of the resolution and should exhibit the same order of magnitude as the black hole's mass, $Q^2 \sim M_\mathrm{BH}^2$. The whole center-of-mass energy $\sqrt{\hat{s}}$ of the parton collision will be captured by the black hole, $\hat{s}=x_1\,x_2\,s=M_\mathrm{BH}^2$, with the production cross section $\hat{\sigma}\!\left(M_\mathrm{BH},\,k\right)$ from above. Since it is irrelevant which parton type forms the black hole and also which momentum the parton has, we sum up/integrate over all possibilities bearing in mind the kinematic constraint above.

In order to capture the properties of the formation process, we are interested in the corresponding differential cross sections. For example, the distribution of the scattering angles at which the produced black holes propagate, is given as the differential cross section with respect to the black holes' momentum fractions $x_\mathrm{F} = x_2-x_1$.
For black hole masses $M_\mathrm{BH} \in \left[M_-,\,M_+\right]$, it turns out to be
\begin{equation}
\frac{\dd \sigma}{\dd x_\mathrm{F}}
= \sum_{a,\,b} \int_{M_-}^{M_+}\!\dd M_\mathrm{BH} \; \frac{2M_\mathrm{BH}}{x_1\,s}\,\,
  f_a\!\left(x_1,\,M_\mathrm{BH}^2\right)\,\,f_b\!\left(\frac{M_\mathrm{BH}^2}{x_1\,s},\,M_\mathrm{BH}^2\right)\,\hat{\sigma}\!\left(M_\mathrm{BH},\,k\right).
\end{equation}
If we want to know the overall mass distribution, we allow the full range of $x_\mathrm{F}$ and calculate
\begin{equation}
\frac{\dd \sigma}{\dd M_\mathrm{BH}}
= \sum_{a,\,b} \int_{0}^{1}\!\dd x_1 \; \frac{2M_\mathrm{BH}}{x_1\,s}\,\,
  f_a\!\left(x_1,\,M_\mathrm{BH}^2\right)\,\,f_b\!\left(\frac{M_\mathrm{BH}^2}{x_1\,s},\,M_\mathrm{BH}^2\right)\,\hat{\sigma}\!\left(M_\mathrm{BH},\,k\right).
\end{equation}
Finally, we find the total production rate of black holes in the accelerator by integrating over the complete mass range and multiplying with the accelerator's luminosity. Consequently, with the LHC design luminosity $L = \SI{e38}{\per\meter\squared\per\second}$ and the total cross section $\sigma = \SI{10}{\nano\barn}$ we expect $\sim \num{e2}$ black holes to be produced per second which scales to $\sim \num{e9}$ black holes per year \cite{BleicherHHS2002,HossenfelderHBS2002}.

In the above ansatz for the cross section, one customarily neglects that there is a minimal mass for black hole formation. In other words, one is tacitly assuming that arbitrary light black holes could be produced as long as the impact parameter is smaller than the corresponding horizon radius. Furthermore, this concept does not consider the fact that at energies around $M_D$ the spacetime departs from its classical description. One of the major consequences is that higher spatial resolutions than a minimal length 
\begin{equation}
l \sim M_D^{-1} = {\left(\frac{8\pi\,M_D^2}{M_\mathrm{Pl}^2}\right)}^{1/k}\,l_\mathrm{c} \ll l_\mathrm{c}
\end{equation}
are no longer possible \cite{Hos12,SNB12}. Both aspects can be taken into account by considering an improved cross section $\hat{\sigma}\!\left(M_\mathrm{BH},\,k\right)$ \cite{MureikaNS2012}:
\begin{equation}
\hat{\sigma}\!\left(M_\mathrm{BH},\,k\right)
= \pi l^2 \, \Gamma\!\left(-1;\,\frac{l^2}{\horizRad^2\!\left(M_\mathrm{BH},\,k\right)}\right) \,
  \theta_l\!\left(M_\mathrm{BH}-M^\mathrm{min}_\mathrm{BH}\right)
\end{equation}
where $\Gamma\!\left(\alpha;\,x\right)$ is the upper incomplete gamma function
\begin{equation}
\Gamma\!\left(\alpha;\,x\right)
\equiv \int_x^\infty\!\dd t \; t^{\alpha-1}\,e^{-t},
\end{equation}
and $\theta_l\!\left(x\right)$ the modified Heaviside step function
\begin{equation}
\theta_l\!\left(x\right)
\equiv \frac{1}{{\left(4\pi\,l^2\right)}^{\frac{1}{2}}} \, 
       \int_{-\infty}^x \!\dd y\; \mathrm{e}^{-\frac{y^2}{4 l^2}}.
\end{equation}
The main advantage of the above cross section is the possibility of capturing a smooth opening of the production channel as well as the implementation of minimal masses consistent with those one can find in a variety of quantum gravity improved black hole metrics \cite{NSS06b,BoR00,MMN11,Mod06,Nic12}.

\section{Life cycle of black holes in particle detectors}

Once the black hole forms, an array of processes is expected to be observed in the particle detector. By considering the semiclassical approximation to quantum gravity, Hawking found out that black holes lose energy: Similar to black bodies they emit thermal radiation, they evaporate, at a temperature $T_\mathrm{BH}$ proportional to their surface gravity $\kappa$ \cite{Haw75} 
\begin{equation}
T_\mathrm{BH}=\frac{\hbar c \kappa}{2\pi k_\mathrm{B}}. 
\end{equation}
The scenario for thermal emission is solid as long as one neglects the quantum backreaction. In other words one assumes that the emitted energy does not affect the background metric.
Based on such hypotheses, one customarily identifies three phases of the black hole life:
\begin{itemize}
\item Balding phase, in which the black hole hair is shed;
\item Spin-down phase, in which the hole moves towards a spherically symmetric configuration through Hawking and Unruh-Starobinskii radiation \cite{Sta73,Unr74};
\item Schwarzschild phase, in which the black hole emits radiation, but in a spherically symmetric way. 
\end{itemize}
Among these three phases, a black hole is assumed to lose most of its mass in the Schwarzschild phase.\footnote{Recent studies have overturned this assumption by showing that the spin down phase characterizes the major part of the black hole life \cite{CDK09}.} After the Schwarzschild phase, the black hole temperature has drastically increased and the semiclassical approximation breaks down. This means that such a terminal phase of the evaporation, called Planck phase, requires a quantum gravity description. In general, it is extremely difficult to extrapolate a consistent scenario for the Planck phase in terms of the current candidate theories to quantum gravity, such as superstring theory and loop quantum gravity. Among other difficulties, one has to face the problem of the lack of a properly defined line element when wild quantum fluctuations affect the spacetime geometry. Furthermore, the semiclassical description is plagued by a  violation of unitarity in quantum mechanics (also known as information paradox), if the black hole evaporated away all its mass prior to an explosive end -- see \cite{GBUP} for some recent discussion.

In the recent years, however, there has been a lot of work aiming to improve the semiclassical approximation by implementing quantum gravity effects in the spacetime of radiating black holes. Among the most notable proposals, we recall here the following quantum gravity improved black hole families, whose name descends from the mechanism employed to improve the classical Schwarzschild metric: noncommutative geometry inspired black hole \cite{Nic05,NSS06a,Nic09,NiS10}, asymptotically safe gravity black holes \cite{BoR00,BoR06}, generalized uncertainty principle black holes \cite{APS01,IMN13,CMN15}, non-local gravity black holes \cite{Nic12,MMN11}, loop quantum black holes \cite{Mod04,Mod06}, quantum $N$ portrait black holes \cite{DvG10,DvG12,DvG13b,FKN16,NiS12} as well as a generic class of non-geometric quantum mechanical models \cite{COY14,NiS14,CMS14,SpS16,SpS16b,SpS15,XCal14}. Interestingly, the metric modifications of almost all the above models converge towards a unique model-independent scenario for the terminal phase of the evaporation. Rather than an explosive end, the black hole undergoes a transition from a negative heat capacity phase to a positive heat capacity phase, called SCRAM phase \cite{Nic09}. Such a scenario was already supported by early investigations of back-reacting black hole spacetimes within perturbative semiclassical techniques \cite{DeS78,BaB88}. At the end of the SCRAM phase the black hole cools down and asymptotically approaches a zero-temperature extremal remnant configuration. Notably such a remnant has been considered as a privilege candidate for a cold dark matter component as originally proposed in \cite{ChA03}. The presence of the SCRAM phase drastically modifies the spectrum of the emitted particles. Such quantum gravity improved black holes tend to emit soft particles mainly on the brane in marked contrast to results based on the Schwarzschild metric \cite{NiW11}. In addition, the presence of the remnant naturally provides the minimal energy for black hole formations, a feature that cannot be captured within purely semiclassical analyses. For further phenomenological repercussions of quantum gravity improved metrics see \cite{Riz06,Gin10}.

\section{Approaches to Experimental Evidence} 

The Hawking emissions we mentioned above are based on the assumption that a black hole evaporates in vacuum and there is no relevant modification of its thermal spectrum. Unfortunately, this is not the case. The thermal emissions in the beginning are the so-called direct emissions. Further effects enter the game afterwards resulting in an effective spectrum that is by far different from the direct one. To capture the subtleties of finally emitted particles from a black hole, we start by providing an analysis of the primary black hole emission along the lines of \cite{CaS06}. 

For microscopic black holes the temperature is of the order of $M_D \sim \SI{1}{\TeV}$ exceeding the rest masses of the Standard Model particles. Furthermore, no gauge interactions are involved, so that the production rates of any Standard Model particle specimens are roughly equivalent. The distribution ${\langle N \rangle}_{\omega s}$ is of a black-body type only depending on the particle's spin $s$ and energy $\omega$,
\begin{equation}
{\langle N \rangle}_{\omega s}
= \frac{{\left|A\right|}^2}{\mathrm{e}^{\omega/T_\mathrm{BH}}-{\left(-1\right)}^{2s}}
\end{equation}
where ${\left|A\right|}^2$ is the greybody factor due to the gravitational potential the particle has to overcome.
However, there are $36$ different types of quarks if one takes into account flavors, colors and antiparticles, and only $6$ types of charged leptons. Therefore the quark production probability is enhanced by a factor of $6$ compared to that of charged leptons. Such and similar considerations are employed to determine the relative particle muliplicity near the horizon of the black hole. 

In order to calculate relative contributions to the radiated power $P_\mathrm{tot}$, the greybody factor is replaced by the cross section $\sigma^{\left(s\right)}\!\left(\omega\right)$ encoding the emission probability of a single particle of spin $s$ and energy $\omega$. The associated emitted power amounts to
\begin{equation}
P
= \frac{1}{2\pi^2} \, \int_0^{\infty}\!\dd\omega\; \omega^3\,\frac{\sigma^{\left(s\right)}\!\left(\omega\right)}{\mathrm{e}^{\omega/T_\mathrm{BH}}-{\left(-1\right)}^{2s}}.
\end{equation}
By summing over the particle multiciplicities one finds that in this so-called direct emission more than 75\% of the emitted power goes back to partons, \textit{i.e.}, quarks and gluons, while photons only carry 1\% to 2\%, depending on the number of extra dimensions. These ratios are in general expected to change with the distance to the black hole because the emitted particles are expected to interact with each other forming a plasma surrounding the event horizon. They can give rise to new particles by hadronization of the partons and by decays. 

If the temperature and density of the emitted particles are high enough, effects like bremsstrahlung and pair production can occur in the QED regime (among electrons and photons) or in the QCD regime (among partons). Each effect increases the number of the particles involved and finally leads to the formation of a plasma: a photosphere (QED) and/or a chromosphere (QCD). The minimum temperature which ensures that on average every emitted particle undergoes at least once such a process is called the critical temperature. Because of the relatively weak coupling constant in QED, the critical temperature for the formation of a photosphere lies at $T_\mathrm{c}^\mathrm{QED} \approx \SI{50}{\GeV}$, while the black hole temperature only has to be above $T_\mathrm{c}^\mathrm{QCD} \approx \SI{175}{\MeV}$ to allow for the chromosphere formation. The existence of these plasmas and the associated interactions shift the energy spectrum of the emitted particles to lower energies so that an observer would effectively observe a lower black hole temperature.

The particle spectrum is supposed to change further: Due to confinement, the newly formed partons will fragmentate into hadrons as soon as their temperature drops below $\Lambda_\mathrm{QCD}$. Some of these hadrons are unstable and decay, \textit{e.g.}, a neutral pion decays into two photons. Both processes enhance the number of other particle types in the final spectrum, especially those of photons, neutrinos, electrons, and positrons.

Two comments about the experimental outcome should be made: First, the results are weakly sensitive to the actual number of extra dimensions as long as $k<5$, since through the Hawking radiation particles are mostly emitted on the 3-dimensional brane. Second, the Hawking radiation could turn out fruitful for particle physics itself: If black holes can be produced in a particle accelerator, their Hawking radiation could exhibit particles whose ordinary production cross sections are extraordinarily small. If those particles are too rare to be analyzed in a particle collision, black holes could provide a more efficient way to create and investigate them.

\section{Black Holes at CERN?}
With a design center-of-mass energy $\sqrt{s}=\SI{14}{\TeV}$, the LHC could either detect microscopic black holes or set stringent constraints on their parameter space. To do so, one needs to define collision properties and observables typical for a black hole formation process. Apart from such creation events also ordinary Standard Model processes could lead to similar signatures. Therefore the Standard Model contribution to the oberservables has to be estimated and subtracted. Significant deviations could hint towards the actual formation of black holes.

In \cite{CMS2012,CMS2011} the CMS group introduced the following scheme: An event is only considered if the sum $S_\mathrm{T}$ of the final state transverse energies amounts to more than \SI{600}{\GeV}, but only jets, leptons, and photons with a transverse energy $E_\mathrm{T}$ higher than \SI{50}{\GeV} are taken into account.\footnote{While longitudinal observables suffer from relics of a particle collision, transverse energy and momentum have newly formed. Therefore the experiment's background is lower.}
In the absence of black holes, the Standard Model predicts QCD multijet events contributing to high values of $S_\mathrm{T}$ only in a small amount due to jets together with a photon/W boson/Z boson, and in an even smaller amount due to the production of (anti) top quarks.

\begin{figure*}[tbp]
\begin{center}
\begin{minipage}{0.45\textwidth}
	\includegraphics[width=\linewidth]{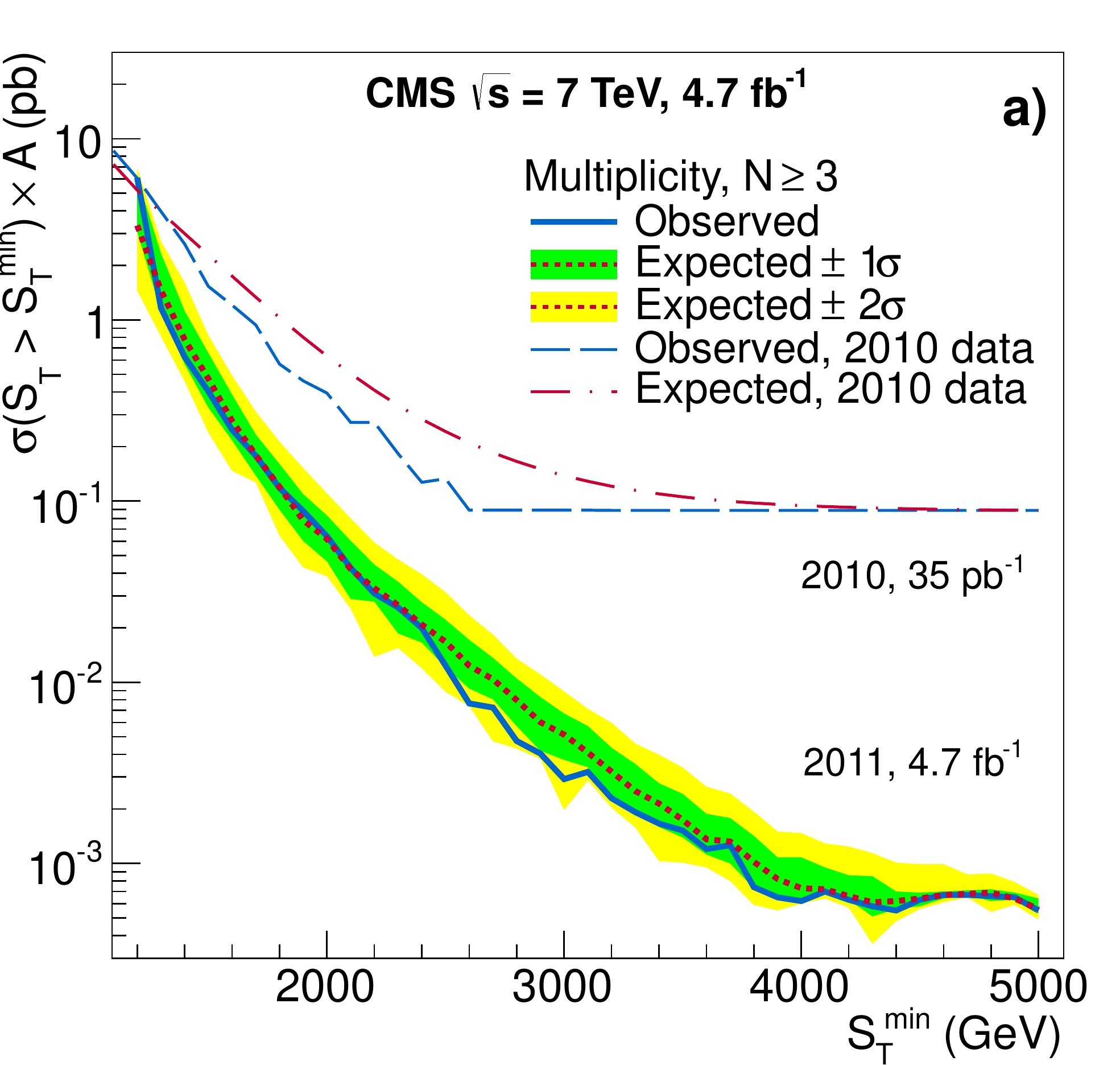}
	\label{fig:CMS_ModelIndep_cross_section}
\end{minipage}
\hspace{0.05\textwidth}
\begin{minipage}{0.45\textwidth}
	\includegraphics[width=\linewidth]{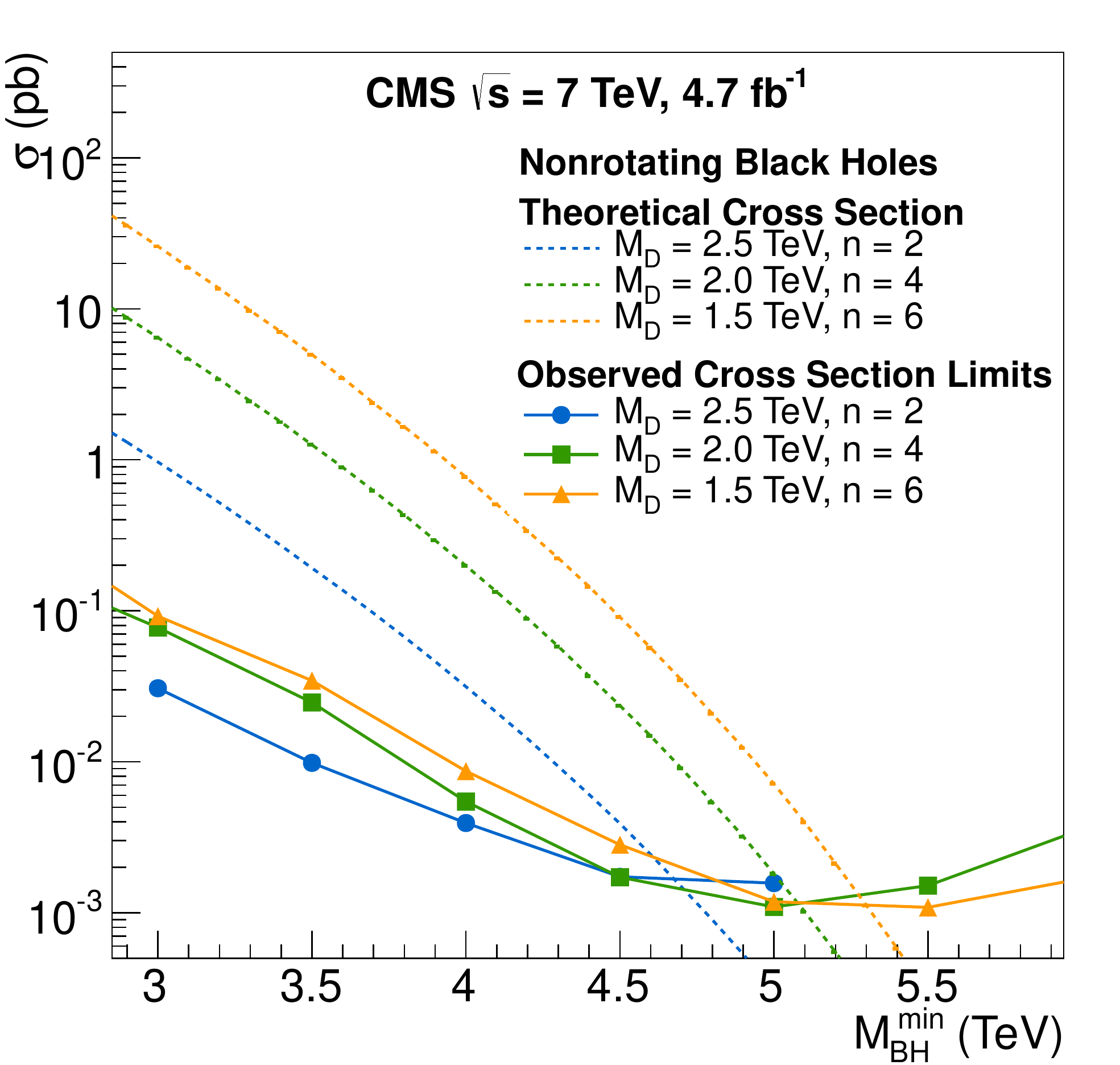}
	\label{fig:CMS_BH_cross_section}
\end{minipage}
\caption{Constraints on cross sections at a confidence level of $95\%$.
	Left: Upper bounds on the cross section $\sigma$ times detector acceptance $A$ for any yet unknown process with multiplicity $N \geq 3$. The constraint depends on the sum $S_\mathrm{T}$ of the final state particles' transverse energies. The graph shows this property by restricting the considered range of $S_\mathrm{T}$ by a lower threshold $S_\mathrm{T}^\mathrm{min}$. The blue solid curve results only from experimental data while the red dotted curve (together with the 1 and 2 standard deviation regions) has been expected, both times a signal acceptance uncertainty of $5\%$ is assumed.
	Right: Experimental constraints on the cross section $\sigma$ of semiclassical black holes in terms the minimal black hole mass $M_\mathrm{BH}^\mathrm{min}$ (solid curves). The theoretical values (dotted curves) are simulated by the BlackMax event generator without remnant formation. The mass range $M_\mathrm{BH}^\mathrm{min} \leq \SI{4.7}{\TeV}$ can be excluded since the theoretical values are larger than allowed by the experimental bounds. The number of extra dimensions is denoted by $n$, the higher-dimensional fundamental mass scale by $M_D$ (cf.~(B.9)).
	Figures from \cite{CMS2012}.}
\label{fig:Cross_Sections}
\end{center}
\end{figure*}

\begin{figure*}[tbp]
\begin{center}
\begin{minipage}{0.45\textwidth}
	\includegraphics[width=\linewidth]{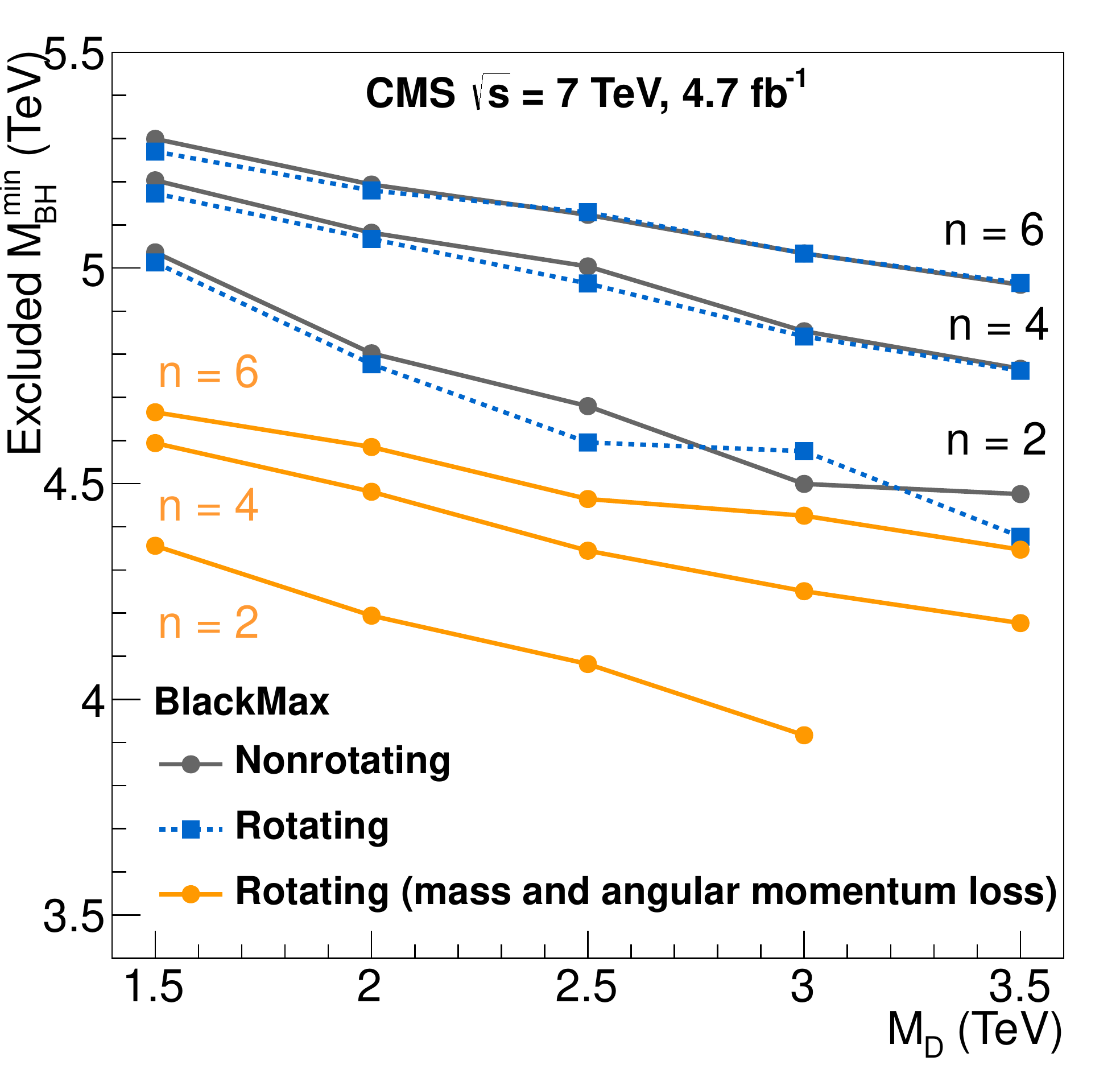}
	\label{fig:CMS_MthMD_Class_BlackMax}
\end{minipage}
\hspace{0.05\textwidth}
\begin{minipage}{0.45\textwidth}
	\includegraphics[width=\linewidth]{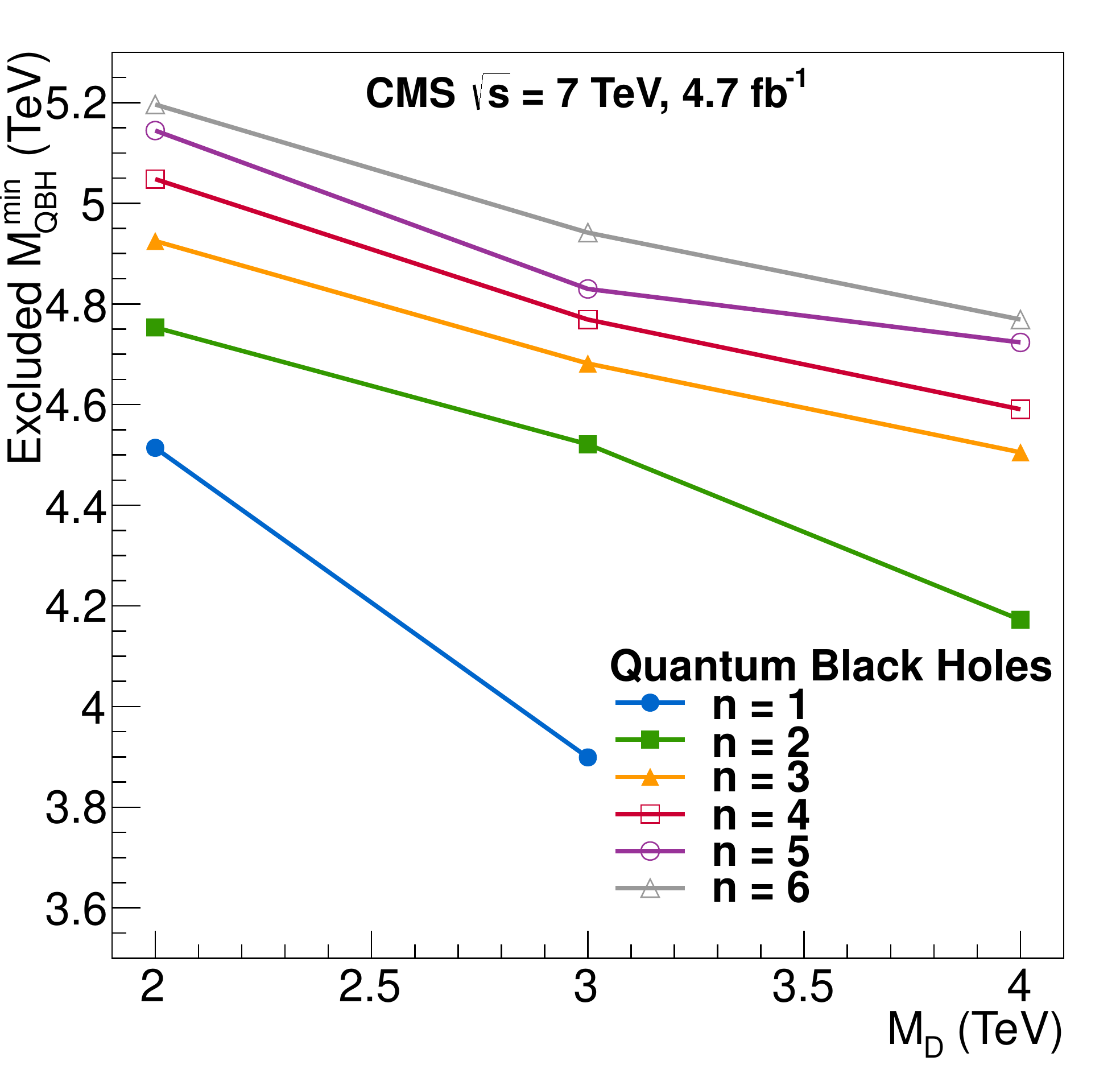}
	\label{fig:CMS_MthMD_Qu_QBH}
\end{minipage}
\caption{95\% confidence level constraints on the minimal black hole mass $M^\mathrm{min}_\mathrm{BH}$ depending on the fundamental energy scale $M_D$ and the number of extra dimensions $n$. 
Left: Semiclassical black holes without final remnant formation based on the BlackMax event generator. This plot explicitly includes the minimal masses which we obtained in the right panel of the preceding figure.
Right: Quantum black holes simulated with the QBH generator \cite{Gingrich2010a,Gingrich2010b}. The curve for $n=1$ is irrelevant since it belongs to another model (Randall-Sundrum).
Figures from \cite{CMS2012}.}
\label{fig:Minimal_Mass}
\end{center}
\end{figure*}

Comparison of experimental data with Standard Model predictions leads to no significant deviation, even if one can still get constraints. This means that, on a general level, the cross section $\sigma$ times the detector acceptance $A$ for any hypothetical process with multiplicity $N \geq 3$ can be constrained. Only taking into account final states with the sum of transverse energy exceeding $S_\mathrm{T}^\mathrm{min}\gtrsim\SI{4.5}{\TeV}$, the aforementioned product reaches approximately $\SI{0.6}{\fb}$ at $95\%$ confidence level (cf.~left panel of Fig.~\ref{fig:Cross_Sections}). On a more specific level, limits on black hole total cross sections can be obtained by employing model-dependent simulations of black hole evaporation scenarios, in our case  the remnant-less BlackMax event generator (cf.~right panel of Fig.~\ref{fig:Cross_Sections}) \cite{DaiSSIRT2008}.
In the small minimal black hole mass range, the theoretical cross sections lie above the corresponding upper limits found by the experiments. Such parameter regions are, therefore, ruled out so that we can derive the minimal black hole mass $M^\mathrm{min}_\mathrm{BH} \geq$ \SIrange[range-phrase = --]{4.7}{5.3}{\TeV} at a confidence level of 95\%. This statement is extended by the analysis of non-rotating black holes represented by the left panel of Fig.~\ref{fig:Minimal_Mass}. Based on semiclassical models one can calculate that the mass is bigger than \SIrange[range-phrase = --]{3.9}{5.3}{\TeV} at $95\%$ confidence level, depending on the number $k \leq 6$ of extra dimensions and the fundamental mass $M_D \leq \SI{4}{\TeV}$. Based on quantum models \cite{Gingrich2010a,Gingrich2010b}  with the same conditions for $k$ and $M_D$, no black hole with a mass lower than \SIrange[range-phrase = --]{4.2}{5.2}{\TeV} is expected to exist at a confidence level of 95\% (cf.~right panel of Fig.~\ref{fig:Minimal_Mass}).

% inserted reference (B.9) by hand!

The latest analysis has been performed by the ATLAS group using Run 2 data at a center-of-mass energy of \SI{13}{\TeV} and integrated luminosity of \SI{3.2}{\per\fm} \cite{ATLAS2016}. The survey selects events with a high-energetic electron or muon and at least 2 other particles or jets, all of which carrying a transverse momentum $p_\mathrm{T}$ of more than \SI{100}{\GeV}. This selection is reasonable since the fraction of charged leptons in the direct emission spectrum amounts to more than $10\%$ \cite{CasanovaS2006}. 

\begin{figure*}[tbp]
  \begin{center}
    \begin{minipage}{0.45\textwidth}
		\includegraphics[width=\linewidth]{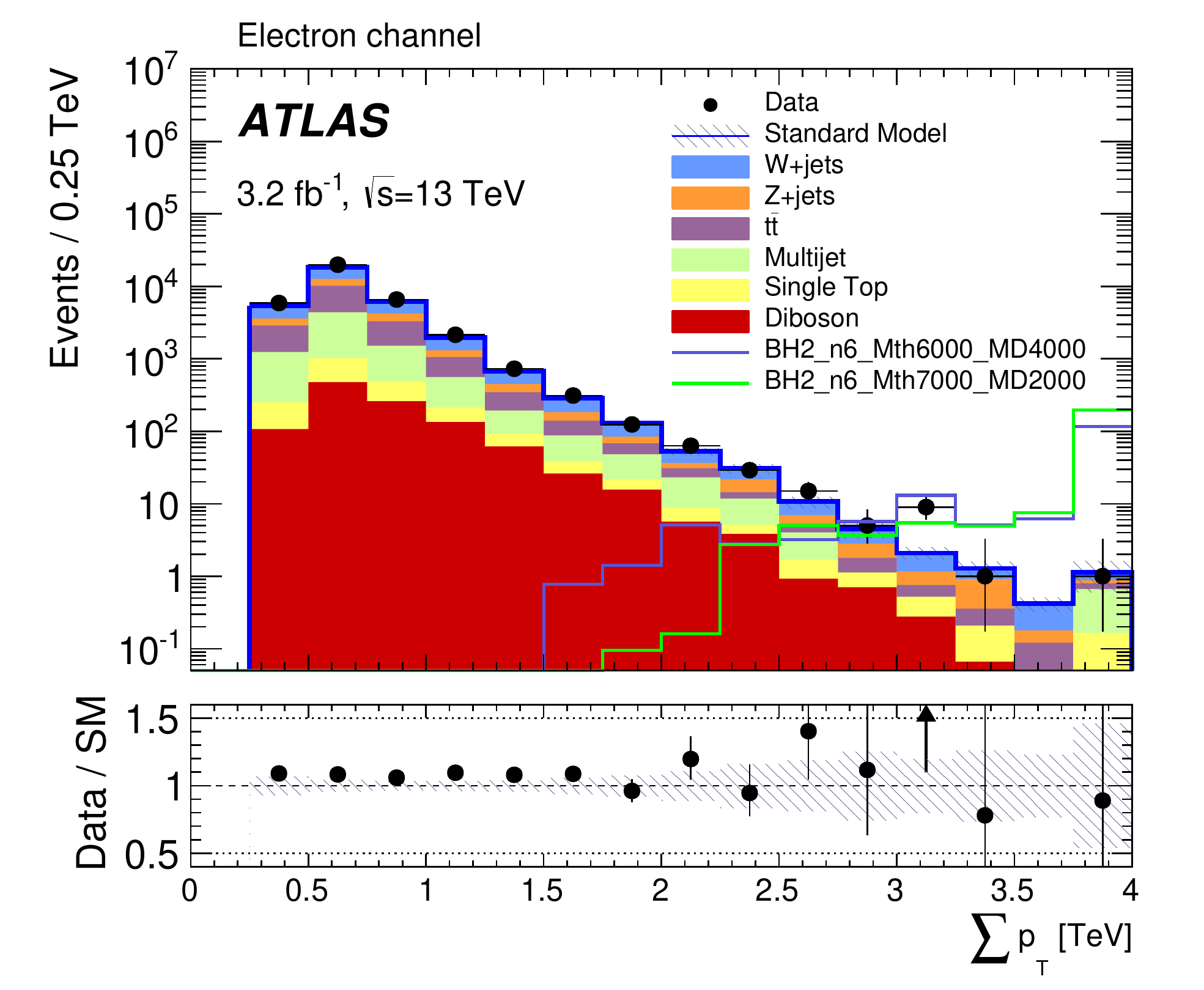}
		\label{fig:ATLAS_Data_SM_Electron}
	\end{minipage}
	\hspace{0.05\textwidth}
	\begin{minipage}{0.45\textwidth}
		\includegraphics[width=\linewidth]{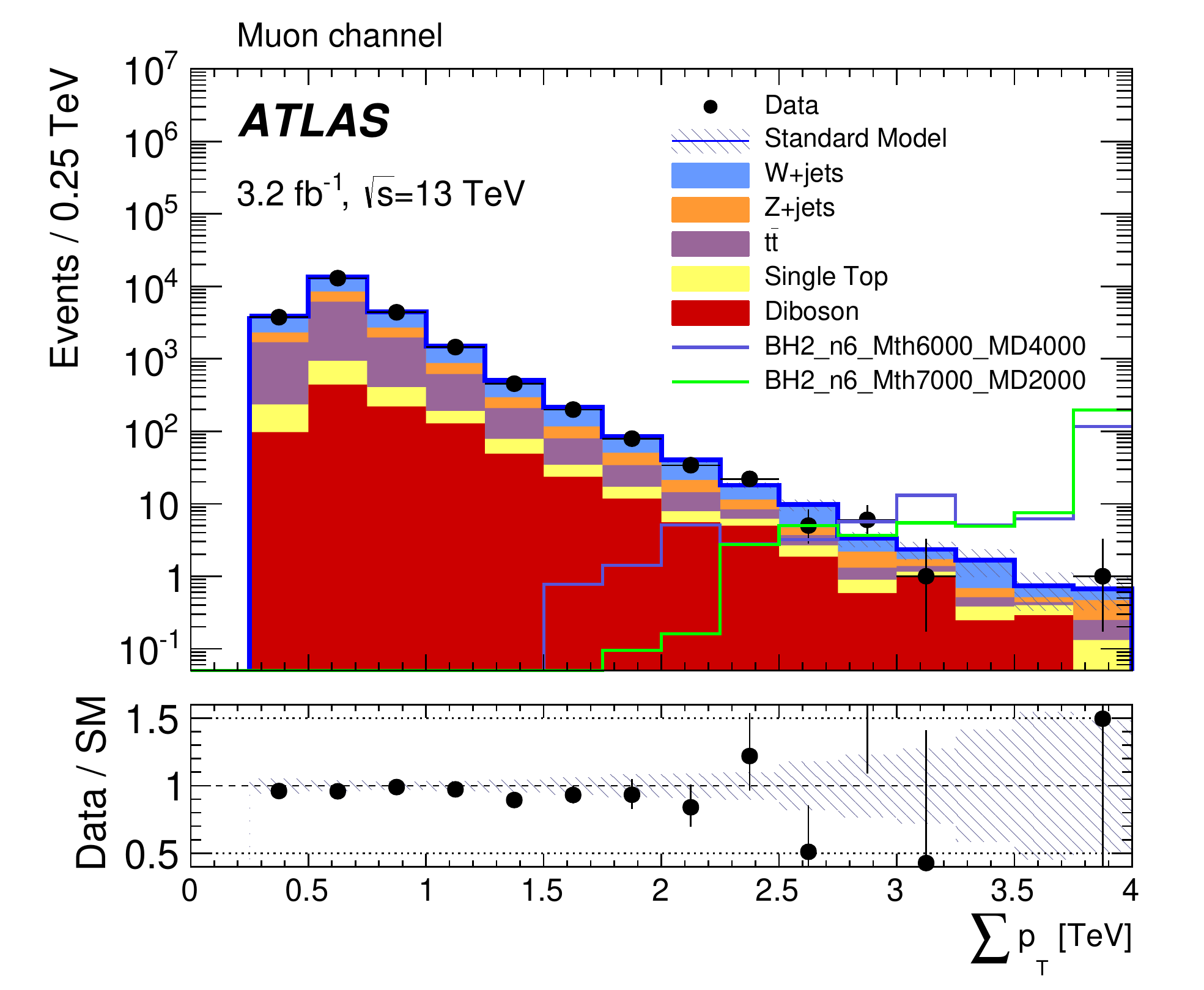}
		\label{fig:ATLAS_Data_SM_Muon}
	\end{minipage}
    \caption{Experimental data vs.~Standard Model predictions. The spectra comprise events whose largest transverse momentum $p_\mathrm{T}$ is carried by an electron (left) or a muon (right). Experiment (black) and the different Standard Model contributions (colored) are presented in the upper panels, their comparison including the Standard Model uncertainties is shown in the lower ones. No significant deviation has been observed. For comparison, sample spectra from rotating black holes in 6 extra dimensions as given by the Charybdis2 Monte-Carlo generator \cite{HarrisRW2003,HarrisK2003,FrostGSCDPW2009} are displayed as the blue and green curves. Figures from \cite{ATLAS2016}.}
    \label{fig:ATLAS_Spectrum}
  \end{center}
\end{figure*}

The events are classified with respect to the sum of all particles' transverse momenta, $\sum p_\mathrm{T}$, and they are taken into consideration if this sum is greater than \SI{2}{\TeV} or \SI{3}{\TeV} respectively. Fig.~\ref{fig:ATLAS_Spectrum} shows the spectra distinguished by whether an electron or a muon carried the highest transverse momentum. For any hypothetical process yielding such final states, the model-independent upper boundary on the cross section times the detector acceptance times its efficiency is \SI{12.1}{\fb} (\SI{2}{\TeV}) or \SI{3.4}{\fb} (\SI{3}{\TeV}), respectively, at a confidence level of 95\%. For rotating black holes in a spacetime with $k=6$ extra dimensions and $M_D=\SI{5}{\TeV}$, the lower mass limit is raised to \SI{7.4}{\TeV}.

\section{Conclusions and Outlook}
The investigation of gravity has undergone a large development now providing predictions accessible by particle accelerators. The hierarchy problem can be solved by the ADD model which introduces compact spacelike extra dimensions accompanied by a new fundamental mass scale for gravity $M_D \geq \SI{1}{\TeV}$. This hypothesis has the chance to be tested in various contexts, \textit{e.g.}, in astronomy by observing the cooling behavior of supernovae. Furthermore, the ADD proposal allows the formation of microscopic black holes in the energy range of particle accelerators, especially of the LHC. 

Up to now, analyses of LHC data do not show significant deviations from the Standard Model predictions caused by microscopic black hole formation, neither semiclassical nor quantum ones. This fact can be used to draw constraints on the cross section and the minimal mass of a black hole: Only black holes with a mass of more than $\sim \SI{4}{\TeV}$ can exist. Measurements and results from Run~2 of the LHC with increased energy and luminosity may provide further insight into the nature of gravity.
The ADD model itself seems to be highly constrained by the latest LHC findings \cite{FGG11}.

On the other hand there is room for a more positive conclusion. Customarily, one employs a semiclassical approach to derive signatures of black hole formation. Such a scenario is valid as long as $M_\mathrm{BH}\gg M_D$ and it breaks down when $M_\mathrm{BH}$ approaches $M_D$ \cite{Par11}. Smaller black holes are purely quantum objects and they might be the result of the SCRAM phase. It is not clear if they keep radiating thermally or if other decaying processes dominate their evolution. In the latter case they might undergo a non-thermal phase characterized by a non-isotropic emission of few high-energetic jets \cite{Cav03}. 
There are other unknowns that are generally neglected when considering black hole signatures. For instance the black hole discharge time may vary drastically if alternative models to ADD are considered like universal extra dimensions. Also the role of color fields \cite{MaW00}, a long lasting spin down phase \cite{CDK09}, the brane tension \cite{KaK06}, the non-stationary nature of the black hole decay, as well as the formation of black hole remnants or a non-geometrical phase in the latest stages of the evaporation are open issues that can potentially modify all the signatures we know so far.

Despite the lack of evidence for black hole formation in particle accelerators, the exclusion still requires a deep analysis and improvement of our knowledge about several crucial features both on the theoretical and on the experimental side.

\newpage

\appendix

\section{The Cavendish-measurable gravitational constant and higher-dimensional gravitational actions}
\label{App_Action}
The Einstein field equations and the Einstein-Hilbert action can be generalized to an arbitrary number of dimensions. Extending the ordinary four dimensional spacetime by $k$ spatial extra dimensions yields the action
\begin{equation}
S
= \frac{1}{16\pi\,\GBareHigher} \, \int\!\mathrm{d}^{4+k}x\;\sqrt{-\hat{g}} \; \hat{R}
  + S_\mathrm{m}\!\left[\hat{g}_{MN};\,\mathrm{matter}\right],
\label{eq:app_higherAction}
\end{equation}
where $\GBareHigher$ denotes the bare gravitational coupling constant in ((3+$k$)+1) dimensions, $\hat{g}$ is the determinant of the higher dimensional metric $\hat{g}_{MN}$ and $\hat{R}=\hat{R}^M_{\;\;M}$ is the corresponding Ricci scalar. The Einstein field equations become
\begin{equation}
\hat{R}_{MN}-\frac{1}{2}\,\hat{R}\,\hat{g}_{MN}
= 8\pi\,\GBareHigher \, \hat{T}_{MN},
\end{equation}
with the Ricci tensor $\hat{R}_{MN}$ and the stress-energy tensor $\hat{T}_{MN}$.

In a 4-dimensional framework, the higher dimensional metric can be decomposed into three kinds of fields which do not possess physical units: the ordinary metric $g_{\mu\nu}$, 4-vector fields $A^a_{\;\;\mu}$, and scalar fields giving rise to the internal metric $G_{mn}$. 
\begin{equation}
\left( \hat{g}_{MN} \right)
= \left(
\begin{array}{cc}
g_{\mu\nu}+A^a_{\;\;\mu}\,A^b_{\;\;\nu}\,G_{ab} & \quad A^a_{\;\;\mu}\,G_{an}\\
A^a_{\;\;\nu}\,G_{am} & \quad G_{mn}
\end{array}
\right).
\end{equation}
Capital Latin indices denote both the ordinary (Greek) and the additional (lowercase Latin) components, \textit{e.g.}, $\left(M\right)=\left(\mu,\,m\right)$ \cite[p.~333]{Ortin2004}.

Now we assume that the extra dimensions are symmetrically compactified to tori, \textit{i.e.}, all compactification radii are identical to $\Rc$. Consistency requires that all fields are periodic in the extra dimensions and thus can be decomposed into Fourier modes based on the period $\lc=2\pi\,\Rc$. We consider the vacuum case and perform a Kaluza-Klein dimensional reduction by restricting ourselves to the zero Fourier modes which are the only massless ones and keep the same notation. Thus there is no dependence on the extra dimensions. The higher dimensional metric determinant $\hat{g}$ and Ricci scalar $\hat{R}$ can now be expressed in 4-dimensional quantities so that we can simply integrate \eqref{eq:app_higherAction} over the extra dimensions:
\begin{eqnarray}
S
&=& \frac{1}{16\pi\,\GBareHigher} \, \int\!\mathrm{d}^{4+k}x\;\sqrt{-g} \; 
    \varphi \left[R-{\left(\partial\,\ln\,\varphi\right)}^2 + \frac{1}{4}\,F^2 
                  - \frac{1}{4}\,\partial_a\,G_{mn}\,\partial^a\,G^{mn}
    \right]\\
&=& \frac{1}{16\pi\,G_{\star}} \, \int\!\mathrm{d}^{4}x\;\sqrt{-g} \; 
    \left[\varphi\,R-\frac{1}{\varphi}\,\partial_\mu\varphi\,\partial^\mu\varphi + \frac{1}{4}\,\varphi\,F^2 - \frac{1}{4}\,\varphi\,\partial_a\,G_{mn}\,\partial^a\,G^{mn}
    \right].
\end{eqnarray}
This is the action in the so-called Jordan frame. Here we introduced the scalar $\varphi$, $\varphi^2 \equiv \left| \det\,G_{mn} \right|$, which scales the volume element of the torus, and $F^2 \equiv F^{m\mu\nu}\,F^n_{\;\;\mu\nu}\,G_{mn}$ where $F^m_{\;\;\mu\nu} \equiv 2 \partial^{}_{[\mu} \, A^m_{\;\;\nu]}$. The index $a$ occurring in the last term is a 4-dimensional frame field Lorentz index \cite[p.~333f.]{Ortin2004}. 
The second line defines the new bare gravitational constant in 4 dimensions, 
\begin{equation}
G_{\star}
= \frac{\GBareHigher}{\lck},
\end{equation}
which in general depends on $\GBareHigher$ and the number of extra dimensions.

We illustrate the relation between the perceivable Newton's gravitational constant $G_\mathrm{N}$ and the bare gravitational constant $G_\star$ in the case of $k=1$ extra dimension below. The corresponding action reads \cite[p.~301]{Ortin2004}
\begin{equation}
S
= \frac{1}{16\pi\,G_{\star}} \, \int\!\mathrm{d}^{4}x\;\sqrt{-g} \; 
  \left[ \varphi\,R-\frac{\varphi^3}{4}\,F_{\mu\nu}\,F^{\mu\nu} \right].
\label{eq:app_fuenfZuVier}
\end{equation}
If we consider the case of vanishing vector fields, the action has the form of a scalar-tensor theory whose general form reads \cite[p.~580]{PeterU2013}
\begin{equation}
S
= \frac{1}{16\pi\,G_{\star}} \, \int\!\mathrm{d}^{4}x\;\sqrt{-g} \; 
  \left[F\!\left(\phi\right)\,R-g^{\mu\nu}\,Z\!\left(\phi\right)\,\partial_\mu\phi\,\partial_\nu\phi - 2U\!\left(\phi\right)\right]
  + S_\mathrm{m}\!\left[g_{\mu\nu};\,\mathrm{matter}\right].
\label{eq:app_ScTensGen}
\end{equation}
In our vacuum case, the coupling function $F\!\left(\phi\right)\equiv \varphi$, the kinetic coefficient $Z\!\left(\phi\right)\equiv 0$, and the potential $U\!\left(\phi\right)\equiv 0$.

The scalar-tensor modified Einstein field equations suggest to define
\begin{equation}
G_\mathrm{eff} \equiv \frac{G_\star}{F\!\left(\phi\right)}
\end{equation}
for the attraction mediated by gravitons. In a Cavendish-like experiment one measures the force $F$ between two test masses $m_1$ and $m_2$ separated by $r$. Both masses couple via the graviton and in addition via the scalar $\phi$. This leads to the definition
\begin{equation}
G_\mathrm{Cav} 
\equiv G_\mathrm{eff}\,\left(1+\alpha^2\right)
= G_\mathrm{eff}\,\left( 
  \frac{2\,Z\!\left(\phi\right)\,F\!\left(\phi\right) 
    + 4\,{\left(\frac{\dd F}{\dd \phi}\right)}^2}
  {2\,Z\!\left(\phi\right)\,F\!\left(\phi\right) 
    + 3\,{\left(\frac{\dd F}{\dd \phi}\right)}^2} 
  \right),
\end{equation}
where $\alpha$ is the coupling constant between matter and the scalar field. The connection with Newton's constant $G_\mathrm{N}$, which we can measure in a Cavendish-like experiment today, 
\begin{equation}
G_\mathrm{N} = \frac{F\,r^2}{m_1\,m_2},
\end{equation}
is given by inserting the appropriate value $\phi_0$ of $\phi$ \cite[p.~582f.]{PeterU2013},
\begin{equation}
G_\mathrm{N} \equiv G_\mathrm{Cav}\!\left(\phi_0\right).
\end{equation}
So in the case of one extra dimension and vanishing vector fields we find
\begin{equation}
G_\mathrm{N} = \frac{4}{3\,\varphi_0} \, G_\star,
\end{equation}
implying that the contribution of the massless scalar in a Cavendish-like experiment amounts to one third of the ordinary graviton contribution.

In general for an arbitrary number of extra dimensions, it is possible to perform a conformal transformation of the 4-dimensional metric accompanied by proper redefinitions of the scalar and vector fields analogously to \cite[p.~304]{Ortin2004}. Indicating the conformal quantities with a tilde, the compactification radius becomes
\begin{equation}
\tilde{R}_\mathrm{c}=\varphi_0\,\Rc
\end{equation}
and the relation between Newton's constant and the conformal bare gravitational constant $\tilde{G}_\star$ for one extra dimension turns out to be
\begin{equation}
G_\mathrm{N} = \frac{4}{3} \, \tilde{G}_\star.
\label{eq:app_NewtFactFive}
\end{equation}

\section{The Cavendish-measurable gravitational constant, the compactified gravitational potential, and the Poisson equation}
\label{App_Poisson}
In regions with a weak field strength we can associate the spacetime with a gravitational potential. In the presence of compact extra dimensions it exhibits a compellingly different behavior for short distances. For distances much smaller than the compactification radius, $r \ll \Rc$, the potential goes like $V \propto r^{-\left(1+k\right)}$ while for $r \gg \Rc$ the potential scales like $V \propto r^{-1}$.

If one zooms into the spacetime structure below the compactification length $l_\mathrm{c}$, the spacetime effectively looks like a flat higher-dimensional one. Thus, we start with considering a flat ((3+$k$)+1)-dimensional Minkowski background metric with $k$ uncompactified extra dimensions. A point mass $M$ in its center produces an isotropic gravitational potential $\potIsoR$ where the radial coordinate $r_{4+k} \equiv \sqrt{\sum_{i=1}^{3+k} {\left(x^i\right)}^2}$ takes into account every spatial dimension. The potential has to fulfill the higher-dimensional Poisson equation
\begin{equation}
\Laplace \potIso = S_{2+k}\,\GPoisson\,\rho_\mathrm{m}
\label{eq:app_HigherPoisson}
\end{equation}
where $\GPoisson$ is the higher dimensional gravitational constant belonging to the (2+$k$)-dimensional unit sphere's surface, $S_{2+k}$. The mass density $\rho_\mathrm{m}$ is chosen to describe the point mass $\rho_\mathrm{m} = M \, \prod_{i=1}^{3+k}\delta\!\left(x^i\right)$. Integration of this differential equation by using Gauss's theorem yields
\begin{equation}
\potIsoR
=-\frac{1}{1+k} \, \frac{\GPoisson\,M}{r_{4+k}^{\;1+k}}.
\label{eq:App_IsoPotential}
\end{equation}
Comparing this result with the derivation in \cite{EmparanR2008} exhibits a relation between $\GPoisson$ and $\GBareHigher$ from \eqref{eq:app_higherAction},
\begin{equation}
\GPoisson
= \frac{4\,\left(1+k\right)\,\Gamma\!\left(\frac{3+k}{2}\right)}{\left(2+k\right)\,\pi^{\left(1+k\right)/2}} \, \GBareHigher
\end{equation}

For $r_{4+k} \gtrsim l_\mathrm{c}$, deviations become relevant. The effects of the symmetric toroidal compactification $x^i \sim x^i+2\pi \Rc,\; 4 \leq i \leq 3+k$ with compactification radius $\Rc$ can easily be analyzed by introducing mirror masses at every point identified with the origin, $x^i=\lc\;n_i$ with $n_i \in \mathbb{Z}$. The usual 3-dimensional radial distance $r \equiv \sqrt{\sum_{i=1}^3 {\left(x^i\right)}^2}$ leads to the exact potential
\begin{equation}
\potr
= -\frac{1}{1+k} 
  \, \frac{\GPoisson\,M}{r^{1+k}} \,
  \sum_{n_4=-\infty}^\infty \dots \sum_{n_{3+k}=-\infty}^\infty\,
  \frac{1}{{\left(1+{\left(\frac{l_\mathrm{c}}{r}\right)}^2 \, 
  \sum_{i=4}^{3+k} n_i^2\right)}^{{\left(1+k\right)}/2}}.
\end{equation}

In the limit $r \gg l_\mathrm{c}$, the product of the sums turns into a $k$-dimensional integral over the total volume of the extra dimensions. After conversion to spherical coordinates one finally ends up with the potential
\begin{equation}
\potr
\simeq - \frac{\pi^{\left(1+k\right)/2}}{2\,\Gamma\!\left(\frac{3+k}{2}\right)}\,\frac{1}{\lck}\,\frac{\GPoisson\,M}{r}
\stackrel{!}{=} -\frac{G_\mathrm{N}\,M}{r}.
\end{equation}
In the last step we inserted the known long-distance Newtonian behavior, which enables us to set up a relation between $\GPoisson$ from the Poisson equation, the bare gravitational constant $G_\star$, and Newton's gravitational constant $G_\mathrm{N}$:
\begin{equation}
\GPoisson
= \frac{4\,\left(1+k\right)\,\Gamma\!\left(\frac{3+k}{2}\right)}{\left(2+k\right)\,\pi^{\left(1+k\right)/2}} \, \lck \, G_\star
= \frac{2\,\Gamma\!\left(\frac{3+k}{2}\right)}{\pi^{\left(1+k\right)/2}} \, \lck \, G_\mathrm{N}
\end{equation}
As pointed out in \cite{Rattazzi2005} the gravitational constant $G_\mathrm{N}$ measured in a Cavendish-like experiment is higher than the ordinary graviton effect due to the scalar contribution.
\begin{equation}
G_\mathrm{N}
= \frac{2\,\left(1+k\right)}{\left(2+k\right)} \, G_\star
> G_\star
\end{equation}
For the case of $k=1$ extra dimension, we consistently obtain the result \eqref{eq:app_NewtFactFive} after the conformal transformation. 

Having discussed the different gravitational constants we are now able to safely generalize the concept of the Planck mass $M_\mathrm{Pl}$. In four dimensions ($k=0$) we set
\begin{equation}
M_\mathrm{Pl} 
\equiv {\left(\frac{1}{G_\star}\right)}^{1/2} 
= {\left(\frac{1}{G_\mathrm{N}}\right)}^{1/2}
\label{eq:App_PlMass}
\end{equation}
based on Newton's and the bare gravitational constant $G_\star$ which coincide in this case. In a $D$-dimensional spacetime with $k$ extra dimensions, $D=4+k$, we define
\begin{equation}
M_D
\equiv {\left( \frac{1}{8\pi}\,\frac{1}{\GBareHigher} \right)}^{1/\left(2+k\right)}
= {\left( \frac{1}{8\pi\,\lck}\,\frac{1}{G_\star} \right)}^{1/\left(2+k\right)}
= {\left( \frac{M_\mathrm{Pl}^2}{8\pi\,\lck} \right)}^{1/\left(2+k\right)}
\label{eq:App_RelationMasses}
\end{equation}
and thus follow the convention of the Particle Data Group \cite{PDG2016} up to conformal rescalings.

\section{Comment on Minimal Black Hole Masses and Associated Radii}
\label{App_Mmin_rmin}
As mentioned in the text we find a minimal mass for black holes using two basic concepts of black hole and quantum physics: If we require that the total mass of a black hole is captured within its event horizon with radius $\horizRad$ and if we define the reduced Compton wavelength $\lambdabarnew_\mathrm{C}$ as the matter's finite extension as usual in quantum mechanics, we obtain:
\begin{eqnarray*}
&& \horizRad \stackrel{!}{=} \lambdabarnew_\mathrm{C} \nonumber\\
&&\longrightarrow \quad 
\begin{array}{l}
M^\mathrm{min}_\mathrm{BH} 
= {\left(\dfrac{\left(2+k\right)\,\pi^{\left(3+k\right)/2}}{\Gamma\!\left(\frac{3+k}{2}\right)}\right)}^{1/\left(2+k\right)} \, M_D \,
= \, {\left(\dfrac{\left(2+k\right)\,\pi^{\left(1+k\right)/2}}{8\,\Gamma\!\left(\frac{3+k}{2}\right)}\right)}^{1/\left(2+k\right)}\,{\left(\dfrac{l_\mathrm{Pl}}{l_\mathrm{c}}\right)}^{k/\left(2+k\right)}\,M_\mathrm{Pl}\\
\\
r^\mathrm{min}_\mathrm{BH}
= {\left(\dfrac{\Gamma\!\left(\frac{3+k}{2}\right)}{\left(2+k\right)\,\pi^{\left(3+k\right)/2}}\right)}^{1/\left(2+k\right)} \, \dfrac{1}{M_D} \,
= \, {\left(\dfrac{8\,\Gamma\!\left(\frac{3+k}{2}\right)}{\left(2+k\right)\,\pi^{\left(1+k\right)/2}}\right)}^{1/\left(2+k\right)}\,{\left(\dfrac{l_\mathrm{c}}{l_\mathrm{Pl}}\right)}^{k/\left(2+k\right)}\,l_\mathrm{Pl}
\end{array}
\label{eq:App_BH_min}
\end{eqnarray*}
This estimate is based on the assumption that both concepts of characteristic length scales are still valid in the quantum gravity regime.

\acknowledgments
This work has been supported by the German Research Foundation (DFG) and the Stiftung Polytechnische Gesellschaft Frankfurt am Main.

% \section*{References}

\bibliographystyle{JHEPforPoS} 
\providecommand{\href}[2]{#2}\begingroup\raggedright\endgroup

% \begin{thebibliography}{99}
% \bibitem{...}
% ....
% \end{thebibliography}

% Questions and answers:
\newpage
%\bigskip
%\bigskip
\noindent {\bf DISCUSSION}

\bigskip
\noindent {\bf GIULIO AURIEMMA:}
I have a comment regarding the CMS results that You have shown, the only thing that we can derive from it is only a lower limit on the mass scale 
%Md 
$M_D$ and on the number of extra dimensions 
%n
$k$, but we cannot exclude the existence of BH with masses much lower than the Planck mass, predicted by the ADD model, simply because they are not expected to have any distinctive signature. In practice is only the fact that the number of events does not exceed the SM. I would like to add also that I have shown in slide 8 of my talk of last Monday, that the present limit after LHC Run II is 
%n>3 
$k>3$ and 
%Md>5600GeV
$M_D>\SI{5600}{\GeV}$.

\bigskip
\noindent {\bf MARCUS BLEICHER:}
Concerning the CMS results, you are right that they are just in terms of exclusion curves. Apart from a recent proposal for sub-Planckian black holes \cite{CMN15}, the Planck scale or the higher dimensional version of it, $M_D$, serves as the minimal mass for a black hole to be formed after a process of matter compression. In any case, the main concerns about experimental signatures of black holes with masses close to $M_D$ are related to the breakdown of the semi-classical approximation in this regime.

\bigskip
\noindent {\bf GENNADY S. BISNOVATYI-KOGAN:}
What are the remnants of the BH evaporation in multi-dimensional gravity?

\bigskip
\noindent {\bf MARCUS BLEICHER:}
Conventional Hawking radiation, which describes black hole evaporation in the semi-classical regime, predicts an increasing black hole temperature for decreasing mass. This implies an explosion at infinite temperature in the end. 

Black hole remnants are proposed final states when the black hole mass reaches the quantum gravity regime, \textit{i.e.}, when it approaches the fundamental mass scale $M_D$. They denote stable configurations at finite temperatures. Two different types have been proposed, hot and cold remnants. The former follow from a semi-classic-like evaporation which stops at a temperature corresponding to the fundamental mass \cite{APS01,ChA03,HossenfelderHBS2002}. The latter species shows a positive specific heat stage in the end which yields remnants at vanishing temperature \cite{NSS06b,Nic09,NiS12,Nic12}. These are typically modeled by extremal black hole configurations.
The main differences among black hole remnants in different numbers of dimensions lie in the quantitative temperature-mass relation and the final mass scale.

% 
% \bigskip
% \noindent {\bf NAME:} Text
% 
%\bigskip
%\noindent {\bf NAME's comment:} Text

\end{document}